\documentclass[pre,twocolumn,floatfix]{revtex4-2}
\setcitestyle{authoryear,round}

\usepackage{amsmath,amsfonts,amssymb}
\usepackage{bm}
\usepackage{mathtools}
\usepackage{graphicx}
\usepackage[english,french]{babel}
\usepackage{chemformula}
\usepackage{soul}
\usepackage[font=small,labelfont=bf,justification=justified]{caption}
\usepackage{subcaption}
\usepackage[T1]{fontenc}
\usepackage{ulem}

\usepackage{siunitx}
\DeclareSIUnit\fps{fps}
\DeclareSIUnit\rpm{rpm}

\usepackage{color}
\newcommand{\correct}{\textcolor{red}}

\begin{document}

\setstcolor{red}


\title{Spatio-temporal boundary dissipation measurement in Taylor-Couette flow using Diffusing-Wave Spectroscopy}
\author{Enzo Francisco}
\affiliation{Universit\'e Paris-Saclay, CNRS, CEA, Service de Physique de l’\'Etat Condens\'e, 91191 Gif-sur-Yvette, France}
\author{Vincent Bouillaut}
\affiliation{ONERA, 29 Av. de la Division Leclerc, BP 72, F-92322 Ch\^atillon Cedex, France}
\author{Tong Wu}
\affiliation{LMFA, Ecole Centrale de Lyon, France}
\author{S\'ebastien Auma\^ \i tre}
\email[Corresponding author. Email address: ]{sebastien.aumaitre@cea.fr}
\affiliation{Universit\'e Paris-Saclay, CNRS, CEA, Service de Physique de l’\'Etat Condens\'e, 91191 Gif-sur-Yvette, France}

\selectlanguage{english}

\begin{abstract}

{Diffusing-Wave Spectroscopy (DWS) allows for the direct measurement of the squared strain-rate tensor. When combined with commonly available high-speed cameras, we show that DWS gives direct access to the spatio-temporal variations of the viscous dissipation rate of a Newtonian fluid flow. The method is demonstrated using a Taylor-Couette (TC) cell filled with a lipid emulsion or a \ch{TiO2} suspension. We image the boundary dissipation rate in a quantitative and time-resolved fashion by shining coherent light at the experimental cell and measuring the local correlation time of the speckle pattern. The results are validated by comparison with the theoretical prediction for an ideal TC flow and with global measurements using a photomultiplier tube and a photon correlator. We illustrate the method by characterizing the spatial organization of the boundary dissipation rate past the Taylor-Couette instability threshold, and its spatio-temporal dynamics in the wavy vortex flow that arises beyond a secondary instability threshold. This study paves the way for direct imaging of the dissipation rate in a large variety of flows, including turbulent ones.}  

\end{abstract}
\maketitle 

\section{Introduction}
\label{Intro}

The determination of velocity gradients yields valuable information in many aspects of fluid dynamics. For instance, they are involved in boundary layer phenomena, drag force and fluid-structures interactions \citep{Guyon2001_Book}. They also play a major role in turbulence theory, where they drive the dissipation and are a key parameter of the theory of wall bounded turbulence \citep{Davidson2015_Book,Robinson1991,Barenblatt1993}.

However, it is difficult to measure them to a sufficient level of spatial and temporal resolution. Indeed, in fluid mechanics, most measurement techniques focus on velocity. Hot wire anemometry gives a temporal evolution of the velocity at a given point with high accuracy \citep{Comte-Bellot1976}.  Nevertheless, to access at least one component of the gradient, one must either assume that Taylor's frozen-flow hypothesis \citep{Frisch1995_Book} holds or add a second wire which might be disturbed by the presence of the first. The estimation remains local and in a single direction. Although less intrusive, Laser Doppler Velocimetry (LDV) is not suitable for gradient measurements either. It remains a local measurement and the temporal resolution is limited by the concentration of seeding particles \citep{Albrecht2002_Book}. Particle Image Velocimetry (PIV) enables imaging of 3 components of the velocity field in a plane. However, the correlation algorithms limit the spatial resolution to about 10 pixels of the camera \citep{Adrian2011_Book}. For instance in the 4th International PIV Challenge, the PIV resolution in the turbulent flow is about a millimeter (case B of \cite{Kahler2016}). Such coarse-graining does not allow for proper derivation of the velocity gradient. This resolution may be improved by zooming in but this reduces the available region of interest. Particle Tracking Velocimetry (PTV) is of no help since the Eulerian resolution is limited by the average distance between tracked particles. To bypass the issues related to particles seeding, one can use Molecular Tagging Velocimetry (MTV) \citep{Gendrich1997}. The fluid displacement is deduced from the grid deformation. With this two-dimensional technique, the gradient resolution is limited by the patterned grid spacing (about 250 $\mu$m in \cite{Gendrich1997}). There are also sensors directly measuring the shear, but they must be placed on a solid surface \citep{Kolitawong2010}. Therefore, they are limited to near-wall boundary layer and are usually local or averaged over the size of the probe.

Our aim here is to present a promising non-intrusive method that allows us to measure quantitatively the norm  of the strain-rate tensor at a boundary :
\begin{equation}
    \frac{\Gamma}{\sqrt{2}}=\sqrt{\sum_{i,j}e_{i,j}^2} 
 \label{defGamma}
\end{equation} 
with a spatial and temporal resolution. $i$ and $j$ stand for the spatial coordinates $\{x,y,z\}$ where $e_{i,j}=\frac{1}{2}(\partial_i v_j + \partial_j v_i)$, $\bm{v}$ being the velocity field. In the case of a pure shear flow, $\Gamma$ reduces to the shear rate. More generally, the energy density dissipation rate by viscosity in a Newtonian fluid is given by $\eta\Gamma ^2$, with $\eta$ the dynamic viscosity of the fluid. Therefore, we are in fact able to obtain a time-dependent 2D map of the dissipation rate at the boundary of a flow. This method, called Diffusing-Wave Spectroscopy (DWS), uses the interfering properties of the coherent light scattered by a turbid fluid.

DWS began to be developed in the late 1980s with the aim of applying the high accuracy of Dynamic Light Scattering spectroscopy to turbid media. It relies on the properties of random light scattering in such turbid media to deduce the average relative displacement of the scatterers \citep{Maret1987,Stephen1988}. The relevance of this approach was first demonstrated by the study of the Brownian motion of the scatterers \citep{Maret1987,Pine1988}. Nowadays, it is commonly used commercially to perform micro-rheology \citep{Mason1997}. Subsequently, the technique was applied experimentally to simple fluid flows \citep{Wu1990,Bicout1994} and studied theoretically for more complex flows \citep{Bicout1991,Bicout1993}. In these pioneering experiments, the dynamics of the scatterers was estimated from the measurement of the decorrelation time of a single far-field speckle. This speckle is selected far from the scattered light source, i.e. the turbid fluid, with a photomultiplier tube (PMT). In that case, DWS gives direct access to $\Gamma$ averaged over the surface, via the intensity fluctuations at the selected speckle following a multiple-scattering process. To speed up the averaging process in the auto-correlation calculation when slow or time-dependent dynamics are at stake, a CCD camera can be used instead of the PMT, to collect the correlation time from several independent far-field speckles and to perform an ensemble average \citep{Viasnoff2002}.

The CCD camera can also be focused on the surface of the flow, for instance on the boundary of a cell. For a given speckle, the backscattered light interfering in this plane is mainly scattered by particles within a surrounding volume of characteristic size $(l^*)^3$, with $l^*$ the transport mean free path (see section \ref{principle}). Fluctuations in the speckle intensity are therefore representative of the scatterers dynamics in the nearby fluid. Thus we can obtain a spatially resolved map of the scatterer dynamics using directly measured local information. This technique was successfully applied mainly in materials science to capture plastic deformations  and their precursors \citep{Erpelding2008,LeBouil2014}. In such studies, in contrast to fluid mechanics, the displacement imposed by the external driving can be as slow as desired. Here we show that thanks to major advances in high-speed cameras, it is now possible to apply this spatially resolved method to hydrodynamic flows. 

In this study, we apply spatially and temporally resolved DWS to the Taylor-Couette (TC) flow extensively studied experimentally and theoretically. This is a necessary step to calibrate the technique and evaluate its limitations. In section \ref{methods}, we detail the principle of the DWS method applied to fluid flow and the conditions required for meaningful measurements. We then describe the experimental setup and detail the procedure to be followed to characterize the optical properties of the fluids and to get reproducible results. The data analysis is also presented. In section \ref{results}, we highlight the agreement between the average of the shear rate tensor norm measured by the camera and by the PMT associated with a photon correlator. These results are also compared to theoretical predictions. The spatial and temporal resolution allows us to observe the Taylor vortices (Taylor vortex flow) and the oscillations of these vortices (wavy vortex flow). The conclusions and perspectives are summarized in section \ref{conclu}.

\section{Measurement method}
\label{methods}
\subsection{Principle of Diffusing-Wave Spectroscopy}
\label{principle}

Details of Diffusing-Wave Spectroscopy (DWS) can be found in \cite{Weitz1993_Chapter,Sheng2006_Chapter,Bicout1993}. We give here the minimal description necessary to apply the method successfully. DWS applies in the multiple-scattering regime, where the transport of light is given by the diffusion approximation. Therefore the photons are supposed to perform a random walk in the turbid medium. The beams (or plane waves) of coherent light scattered by the turbid medium interfere and lead to a speckle pattern sparkling with time. The  speckle pattern depends on the geometry, but the light decorrelation of a given speckle traces back the dynamics of the scatterers in the fluid domain explored by the interfering beams. More precisely, the light decorrelation at a given point $\bm{r}$ outside the fluid is characterized by the correlation function of the electric field $g_1$ : 
\begin{equation}
    g_1(\tau)=\frac{\langle \bm{E}(\bm{r},t)\cdot\bm{E^*}(\bm{r},t+\tau)\rangle }{\left[\langle\lvert \bm{E}(\bm{r},t)\rvert^2 \rangle\langle\lvert \bm{E}(\bm{r},t+\tau)\rvert^2 \rangle\right]^{1/2}}
\end{equation}
where $\bm{E}$ is the complex electric field, $\langle ~\cdot~\rangle$ corresponds to an ensemble average and $~\cdot~^*$ defines the complex conjugate. In the following we will only consider quasi-stationary processes, therefore the denominator can be replaced by $\langle\lvert \bm{E}(\tau)(\bm{r},t)\rvert^2 \rangle$ and the averaging can be done over time $t$. Actually one can only access to the correlation function of the light intensity :
\begin{equation}
 g_2(\tau)=\frac{\langle I(\bm{r},t)I(\bm{r},t+\tau)\rangle }{\langle I(\bm{r},t)\rangle^2}
 \label{g2}
 \end{equation}
 where $I(\bm{r},t)= \bm{E}(\bm{r},t)\cdot\bm{E^*} (\bm{r},t)=\lvert \bm{E}(\bm{r},t) \rvert^2$. However, as we average over a large number of independent scattering events, one can show that $g_2$ is related to $g_1$ by the Siegert relation \citep{Ferreira2020}:
 \begin{equation}
 g_2(\tau)=1+\beta |g_1(\tau)|^2 
\label{Siegert}
\end{equation}
with $\beta=\frac{\langle I(\bm{r},t)^2\rangle - \langle I(\bm{r},t)\rangle^2 }{\langle I(\bm{r},t)\rangle^2}$ the contrast, which can be up to 1 in our case. Since $g_1$ decreases from 1 (full correlation) at $\tau=0$ to $0$ (full decorrelation) at $\tau\rightarrow\infty$, $\beta$ is given by $\beta=g_2(0)-1$. Therefore $g_1$ can be deduced directly from the measurement of $g_2$.

In nearly all cases of practical interest, the scattering from each particle is weak enough to neglect localization and coherent effects but also to approximate the scattered waves by plane waves (Born approximation). Then $g_1$ can be expressed as a sum over path lengths \citep{Pine1988}:
\begin{equation}
g_1(\tau)=\int_0^\infty P(s) \langle \exp\left(i\Delta \Phi_s(\tau)\right)\rangle_s ds
\label{g1}
\end{equation}
where $\Delta \Phi_s(\tau)$ is the phase shift of the light due to the scatterers displacement along a given optical path of length $s$. $\langle ~\cdot~\rangle_s$ is an average over all the  optical paths of length $s$ and $P(s)$ is the probability to get a path of length $s$. The probability $P(s)$ can be deduced directly from the diffusion theory for a given geometry. Indeed, in the diffusion approximation, the photons perform a random walk with a mean free path $l$. The mean free path is given by $l=1/(\sigma c)$ where $c$ is the number of scatterers per unit volume and $\sigma$ is the scattering cross section, which depends on the scatterer considered and the wavelength. However, the scattering may be anisotropic for large enough particles. Therefore we have to introduce the transport mean free path $l^*=l/\left(1-\langle \cos{\theta}\rangle \right)$, with $\theta$ the angle between the scattered wave vector and the incident wave vector and $\langle ~\cdot~ \rangle$ an averaging over many scattering events. The transport mean free path is the distance a photon must travel before its direction is randomized. In the multiple-scattering regime, the photons therefore perform an isotropic random walk with a mean free path $l^*$.

All the information about the dynamics of the scatterers is contained in the phase shift $\Delta \Phi_s(\tau)$. It can be written as $\Delta\Phi_s(\tau) = \sum_{i=1}^{n} \bm{q_i}\cdot\bm{\Delta r_i(\tau)}$, where $\bm{q_i} = \bm{k_i} - \bm{k_{i-1}}$ is the scattering wave vector, i.e. the difference between the wave vectors before and after the $i^{th}$ scattering event, and $\bm{\Delta r_i(\tau)}$ is the displacement of the $i^{th}$ scatterer during time $\tau$. The number of scattering events in the considered path is $n \approx s/l^*$ in the diffusion approximation. In the multiple-scattering regime, $\Delta \Phi_s(\tau)$ is the sum of independent phase shifts induced by independent scattering events. We can therefore apply the central limit theorem
to this sum of independent events and expect a Gaussian
distribution of the phase shift $\Delta \Phi_s(\tau)$, such that:
\begin{eqnarray}
\langle \exp\left(i\Delta \Phi_s(\tau)\right)\rangle_s=& \exp\langle i\Delta \Phi_s(\tau)\rangle_s\cdot \nonumber\\
&\exp\left(-\frac{\langle {\Delta \Phi_s}^2(\tau)\rangle_s-\langle \Delta \Phi_s(\tau)\rangle_s^2}{
2}\right)
\label{avPhi}
\end{eqnarray}
Hence the two first moments of $\Delta \Phi_s(\tau)$ encompass the whole dynamics. 

The computation of these moments depends on the specific problem under consideration. The simplest case is a medium at rest, so the scatterers only undergo Brownian motion. In that case, one can show that $\langle \Delta \Phi_s(\tau)\rangle_s=0$ and $\langle {\Delta \Phi_s}^2(\tau)\rangle_s=4Dk^2\tau s/l^*$ with $D$ the diffusion coefficient of the particles \citep{Maret1987,Pine1988}. If only a fluid flow is at play, as long as  the smallest characteristic length scale of the flow $\Lambda$  is much larger than $l^*$, one can develop the relative displacement of the scatterers into a 1$^{st}$ order Tayor expansion. We also consider small $\tau$ compared to the characteristic evolution time of the flow in order to assume a ballistic displacement of the scatterers. Under these conditions,  $\langle \Delta \Phi_s(\tau)\rangle_s=0$ in incompressible flows because it is proportional to the velocity divergence. Moreover, one can show that $\langle {\Delta \Phi_s}^2(\tau)\rangle_s=4 \left( \frac{\tilde{\Gamma}(s) l^* k }{\sqrt{30}} \right)^2 \tau^2 s/l^*$ \citep{Bicout1993,Wu1990}, where : 

\begin{equation}
    \tilde{\Gamma}(s)=\sqrt{2\left\langle\sum_{i,j}  e_{i,j}^2 \right\rangle_s}
    \label{Eq:Gamma}
\end{equation}
Actually the dependence over the path length $s$ can be dropped ($\tilde{\Gamma}(s)=\Gamma$) as long as the velocity gradients do not strongly evolve along a path, which is ensured if $l^*\ll \Lambda$ \citep{Erpelding2010}.

In our experiments, both contributions from the Brownian motion and the flow have to be taken into account. Since we consider a ballistic displacement of the scatterers regarding the flow, the phase shift is simply given by the sum of the diffusive contribution (the Brownian motion) and the convective contribution (the flow), which are independent. In the end, only the 2$^{nd}$ moment is non-zero and it is given by :
\begin{equation}
    \langle {\Delta \Phi_s}^2(\tau)\rangle_s=4 \frac{\tau}{\tau_0} \frac{s}{l^*}+4 \frac{\tau^2}{\tau_v^2} \frac{s}{l^*}
    \label{Eq:PhaseFluctuations}
\end{equation} 
where $\tau_o=1/(Dk^2)$ is the characteristic correlation time induced by the Brownian motion and $\tau_v=\sqrt{30}/(\Gamma kl^*)$ is the characteristic correlation time due to the velocity gradient.

Because $P(s)$ depends on the geometry, so does the precise shape of the function $g_1(\tau)$. Several examples have been computed exactly \citep{Weitz1993_Chapter,Bicout1991} (see also appendix \ref{Appendix1}). For the backscattering geometry with uniform illumination of the incident face, in the limit of a semi-infinite system, it simply decays exponentially :

\begin{equation}
g_1(\tau)\approx\exp\left(-\gamma\sqrt{6\left [\tau/\tau_o+(\tau/\tau_v)^2\right]}\right)
\label{g1exp}
\end{equation}
where $\gamma l^*$ can be interpreted as an effective distance necessary for non diffusive incident light to become diffusive inside the sample (see appendix \ref{Appendix1}). The parameter $\gamma$ takes into account the reflections at the boundaries and depends on several parameters : the geometry of the cell, the refractive indices of the fluid and the cell and the presence of a polarizer or analyzer \citep{MacKintosh1989,Zhu1991}. However, it can be  determined {\it in situ} by studying the Brownian motion of the fluid in the cell without flow (see section \ref{gamma}).

 We know from the diffusion approximation that the fluid volume probed by the backscattered light remains confined in the vicinity of the incident surface, i.e. in a small layer of thickness of a few $l^*$. Since the thickness $L$ of the cell (the gap $L$ in the TC flow) is much greater than $l^*$, we will consider that $\Gamma$ is probed at the incident surface (the boundary between the outer cylinder and the fluid in the TC flow). In the same spirit, when the high-speed camera is focused on this surface, the intensity at a certain pixel comes from interfering beams that have most probably explored a volume of a few ${l^*}^3$  \citep{Erpelding2008}. Since the camera pixel is larger than $l^*$, $\Gamma$ is probed at the surface on a pixel-sized area. Consequently, by considering the light intensity decorrelation of each pixel, we can measure the local norm of the strain-rate tensor at the surface, $\Gamma(y,z)$.

It is therefore possible to probe $\Gamma$ with DWS as long as the Brownian motion correlation time $\tau_0$, the dimensionless coefficient $\gamma$ and the transport mean free path of the light in the turbid media $l^*$ are previously determined. The proper interpretation of the data also requires the following conditions :

\begin{itemize}
\item The scattering from each particle has to be weak enough to neglect localization and coherent effects. A scattered wave also has to be approximated as a plane wave when it reaches the next scatterer. Therefore we need the mean free path to be much greater than the wavelength $\lambda$ of the light in the medium : $l \gg \lambda$.
\item The multiple-scattering regime requires that many scattering events can occur and therefore that the thickness of the cell is much greater than the transport mean free path : $L\gg l^*$ where $L$ is the characteristic size of the system. This also justifies the semi-infinite approximation in the backscattering geometry.
\item $\Gamma$ will be properly probed if and only if the smallest characteristic length scale of the flow is much larger than the transport mean free path : $\Lambda \gg l^*$. This also ensures that we observe the exponential decay of equation \eqref{g1exp}.
\item The ballistic displacement of the scatterer is ensured if the correlation time due to the velocity gradient is much smaller than any characteristic evolution time of the flow.
\item The concentration of scatterers must be uniform within the fluid in order to get a uniform transport mean free path of the light $l^*$ in the turbid media.
\end{itemize}

It is important to notice that with this technique, the  proper estimation of the velocity gradient is not limited by the camera spatial resolution since we measure directly $\Gamma$ instead of deriving it from a measurement of the velocity.  This proper estimationis only limited by $l^*$ which is controlled by the particle concentration, while the pixel field of view gives the area over which $\Gamma$ is averaged. In contrast with PIV measurements where the gradient estimate depends on the camera resolution, here this area can simply be adjusted as required by zooming in or out, depending on the total region of interest, the number of pixels and the size of the studied structures. However, the camera or the PMT still needs to be fast enough to accurately measure the decay of the intensity autocorrelation.

\subsection{Experimental setup}
\label{Setup}

\subsubsection{Taylor-Couette flow}
\label{TCFlow}

In order to benchmark the DWS method, we apply it to the well-known TC flow. Indeed, this flow is one of the paradigmatic systems of fluid mechanics. The first instabilities have been widely reported in many publications \citep{Andereck1986}. A convenient control parameter of the instability is the Taylor number : $Ta=\Omega^2 L^3 R_i/\nu^2$ where $\Omega$ is the rotation rate of the inner cylinder (in rad/s), $L=R_o-R_i$ is the fluid gap between the outer cylinder (of radius $R_o$) and the inner cylinder (of radius $R_i$) and $\nu$ is the kinematic viscosity. The laminar base flow (circular Couette regime) is a pure shear flow which can be computed exactly for an infinitely long cell \citep{Guyon2001_Book}. The shear rate (and therefore $\Gamma$) at a radius $r$ ($R_i \leq r \leq R_o$) is then : $\Gamma_{th}(r)= 2 \Omega R_i^2R_o^2/[r^2 (R_o^2-R_i^2)]$. The first instability (Taylor vortex regime) generates steady rolls called Taylor vortices at $Ta\geq Ta_c\approx1712$ (the exact threshold actually depends on the radius ratio, see for instance \cite{DiPrima1984}). By increasing $Ta$ further, one can observe an unsteady wavy instability of the vortices (wavy vortex regime). DWS has already been applied to this flow in the nineties \citep{Bicout1994}. In this pioneering work, by using a PMT and a correlator, the authors performed global measurements with an extended plane wave and point-like measurements using a single beam with a waist of 1 mm. The main outcome of our work is to go further by the use of a high-speed camera to get a full time-dependent map of the norm of the strain-rate tensor at the boundary of the Taylor-Couette flow.    

We test the method using two turbid fluids. The choice of these two turbid fluids is driven by their properties (stability, viscoelasticity, surface tension, cost, ...). The first one is a suspension of titanium dioxide (\ch{TiO2}) anatase particles from Kronos 1002, in deionized water, with a concentration of \SI{10}{\gram/\liter}. The second one, called Intralipid 20\%, is a stabilized lipid emulsion made of drops of soybean oil suspended in water, with a volume concentration of 20\%. Our TC flow is generated in two different cells, adapted to each fluid viscosity. They are designed so that a clear linear regime (for $\Gamma$) and the first instabilities can be observed in our range of accessible parameters. Both of them are made of two coaxial Plexiglas cylinders of height $H=\SI{10}{\cm}$. The first cell, used for the $\mathrm{TiO_2}$ suspension, has an inner radius $R_i=\SI{12.9}{\mm}$ and an outer radius $R_o=\SI{15}{\mm}$, therefore a gap $L=\SI{2.1}{\mm}$. The outside of the outer cylinder was shaped as a plane facing the camera to eliminate optical aberrations. The second cell, used for the lipid emulsion, is slightly larger since its viscosity is higher (see section \ref{viscosity}) : $R_i=\SI{12.9}{\mm}$, $R_o=\SI{15.75}{\mm}$ and $L=\SI{2.85}{\mm}$. In this case, the outer cylinder is immersed in a square tank filled with clear water, also in order to reduce the optical aberrations. For both cells, the outer cylinder is fixed and the inner cylinder is driven at a given rotation rate with a rheometer head (Anton Paar SDR301). The rheometer gives a precise measurement of the torque applied to the inner cylinder, which enables us to determine the fluid's viscosity (see \ref{viscosity}).

\subsubsection{Optical arrangement and measuring systems}
\label{Optics}

\begin{figure}[h!]
\includegraphics[width=1\columnwidth,angle=0]{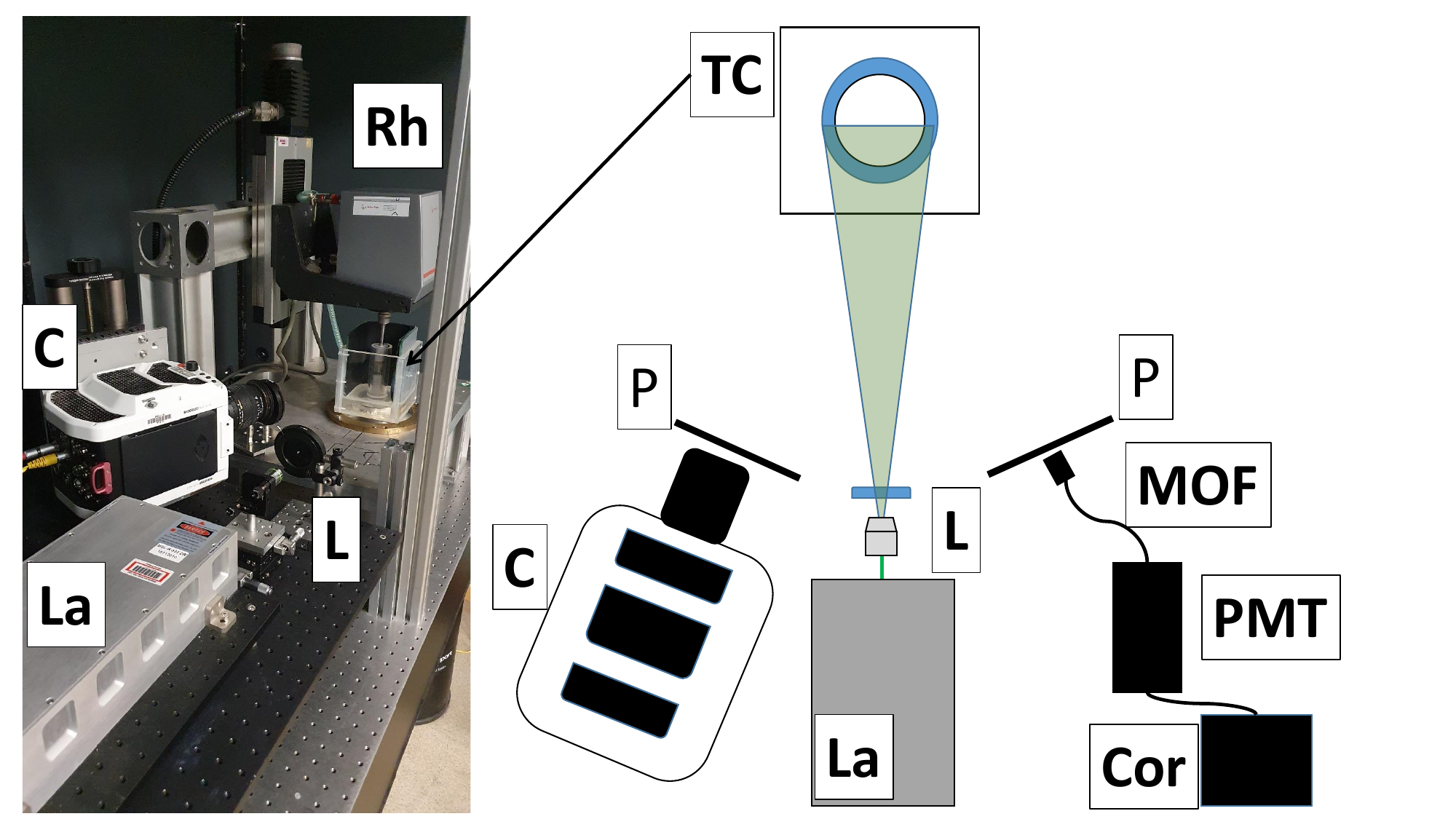} 
\caption{{\bf Left} Picture and {\bf Right} sketch of the experimental setup. A coherent beam produced by the laser ({\bf La}) is enlarged by a microscopic lens (X20) and elongated by a cylindrical lens ({\bf L}) in order to illuminate the entire front of a Taylor-Couette cell ({\bf TC}) filled with turbid fluid. The backscattered light is collected through a polarizer ({\bf P}) by the high-speed camera ({\bf C}) focused on the cell. A far-field speckle of backscattered light is also selected through a polarizer ({\bf P}) by a monomode optical fiber ({\bf MOF}) linked to a photomultiplier tube ({\bf PMT}) and a correlator ({\bf Cor}) (not shown in the picture). The inner cylinder of the Taylor-Couette cell is driven at constant rotation speed by a rheometer head ({\bf Rh}) (not depicted in the sketch).} 
\label{DWSTopview} 
\end{figure}

The optical arrangement, depicted on Figure \ref{DWSTopview}, includes a polarised laser source  (CNI model MSL-R-532-2000) of power \SI{2}{\W} and wavelength $\lambda=\SI{532}{\nm}$. The laser beam is enlarged by a microscope lens (X20) and elongated by a cylindrical lens in order to illuminate uniformly the entire cylinder. The backscattered light is collected by a photomultiplier tube (PMT Hamamatsu H9305-04) through a single-mode optical fiber. The monomode fiber is necessary to ensure a speckle-like selection \citep{Brown1987}. Its numerical aperture is about 0.1 to 0.14 and it is located at \SI{20}{\cm} from the boundary between the fluid and the outer cylinder. Therefore, the backscattered light is collected from a disk of radius 2 to \SI{2.8}{\cm}. A photon correlator (FLEX02-01D) is connected to the PMT to directly compute the correlation of the light intensity, with an acquisition time of \SI{1.28}{\micro\s}. This part of the setup allows us to recover the results previously obtained on this flow \citep{Bicout1994}.

An ultra high-speed camera Phantom V2010 is added to get the spatial resolution. The camera focuses on the boundary between the outer cylinder and the flow and thus captures the photons just escaping from the cell, from several near-field speckles. The correlation time of these speckles (and therefore of the corresponding pixel) probes the local velocity gradient. By doing so, we can image $\Gamma$ over an area of 64$\times$128 pixels (Width$\times$Height). Depending on the level of zoom we choose, the pixel size varies but it is about \SI{250}{\micro\m}, so the measurement surface is about $\SI{1.6}{\cm}$$\times$$\SI{3.2}{\cm}$. Indeed, the frame rate of this model is up to 22 600 frames per second (fps) in full resolution but we reduce the resolution to reach \SI{400 000}{\fps}, corresponding to an acquisition time of $\Delta t=\SI{2.5}{\micro\s}$. The characteristic decay time measured when the decorrelation is dominated by the velocity gradient is given by equation \eqref{g1exp} : $\tau_{meas}=\tau_v/(\sqrt{6}\gamma)=\sqrt{5}/(\gamma k l^*\Gamma)$. In our case, it is of order : $\tau_{meas}\approx2.10^{-3}/\Gamma$. Thus, the camera is fast enough as long as the shear rate is not too strong : if $\Gamma\leq\SI{200}{\per\s}$, then $\tau_{meas}\geq4\Delta t$ which is enough to correctly fit $\tau_{meas}$. Note that $\Gamma$ can locally be substantially larger than the linear estimation ($\Gamma=\Omega R_i /L$) as soon as the Taylor vortices are at play. Therefore we were not able to perform the experiment beyond a maximum rotation rate of \SI{100}{\rpm}.

At such acquisition frequency, the light intensity on the CCD sensor of the camera is the main issue. This is why we do not use any diaphragm to control the speckle size. Indeed, inserting a diaphragm does not significantly increase the contrast $\beta$, because the intensity drops and becomes too low in comparison to the CCD sensor sensibility. Typically, the speckle pattern for a given pixel exhibits a contrast of about 0.8\%, which is enough to get the right correlation function (see Figure \ref{SelfCorr}). 

Two polarizers, cross-polarized with the laser, are put in front of the camera and the fiber aperture to remove specular reflection. However, a scattering event can only slightly modify the polarization of the incident wave. Hence, when only cross-polarized light is collected, the contribution from short paths in the path distribution $P(s)$ is reduced while the contribution from long paths, which have completely randomized the polarization, is enhanced. This effect has been widely shown to influence only the value of $\gamma$ without violating the DWS theory \citep{Pine1990,Weitz1993_Chapter}.

\subsection{Experimental procedure}
\label{procedure}

To probe $\Gamma$ using DWS, we first have to determine the Brownian motion correlation time $\tau_0$, the dimensionless coefficient $\gamma$ and the transport mean free path of the light in the turbid media $l^*$. We also measure the kinematic viscosity $\nu$ of the turbid fluids to compute the Taylor number.

\subsubsection{Fluid preparation and determination of $\tau_0$}
\label{fluid tau0}

The \ch{TiO2} powder is dispersed in deionized water with a concentration of \SI{10}{\gram/\liter}. To compute the Brownian motion correlation time $\tau_0=1/(k^2D)$, one needs to determine the mean particle diameter $a$. Indeed, according to the Stokes-Einstein formula, the diffusion coefficient $D$ is given by $D=\frac{k_BT}{6\pi\eta}(a/2)$  where $T$ is the temperature, $k_B$ is the Boltzmann constant, $\eta$ is the dynamic viscosity of the fluid carrying the scatterers (water in both of our fluid suspensions) and $a$ the mean particle diameter. To determine the particle size, we used the Zetasizer from Malvern Panalytical (Ver. 7.03) based on Dynamic Light Spectroscopy. A mean particle diameter $a_{TiO_2}=\SI{0.4}{\micro\meter}$ was found, in agreement with the provider's data and with the others techniques we used like Atomic Force Microscopy and Scanning Electron Microscopy. However, the \ch{TiO2} particles tend to flock. By increasing the mean particle diameter, flocking has 3 detrimental effects : it changes the Brownian motion correlation time, alters the optical properties (in particular, the transport mean free path), and enhances sedimentation making the system inhomogeneous. Indeed, the \ch{TiO2} anatase has a relative density around 3.8 and therefore tends to sediment in water. Sedimentation has to be as limited as possible, because we need the scatterers to be uniformly dispersed in the fluid with a known concentration to get a uniform transport mean free path in the fluid (see section \ref{l*}). To prevent as much as possible the particles from flocking, we disperse them in deionised water (with a resistivity of \SI{18}{\mega\ohm\cm}) and we immerse our sample in an ultra-sonic bath for 5 minutes to break up any clusters of particles. By following this procedure, we are able to maintain a stable suspension for several hours (the $\gamma$ parameter, which captures some optical property of the suspension, changes by only 5\% in 24 hours) while our measurement run lasts only one hour. However, this procedure may not be applicable in different installations of larger volume or when the fluid is difficult to fill in.

This is why we also used a stabilised emulsion of Intralipid 20\%. As it is stabilized, this lipid emulsion is not affected by flocking or sedimentation. However, it is expensive and must be conserved at low temperature, whereas the \ch{TiO2} powder is cheap and can be stored at room temperature. Care should be taken to ensure that the sample of Intralipid 20\% thermalises at room temperature before use. The mean diameter of the scatterers (the drops of soybean oil) was also obtained with DLS measurements : $a_{lipid}=\SI{0.28}{\micro\meter}$. Hence, the following values of the Brownian motion correlation time $\tau_0$ at 24°C are obtained : $\tau_0=$\SI{3.39e-3}{\s} for the \ch{TiO2} suspension and $\tau_0=$\SI{2.37e-3}{\s} for the lipid emulsion (see Table \ref{Tab:VariablesFluids}).

\subsubsection{Determination of $\gamma$}
\label{gamma}

The dimensionless parameter $\gamma$ is linked to the boundary conditions chosen to solve the diffusion equation (see appendix \ref{Appendix1}). It depends on the geometry of the cell, the refractive indices of the fluid and the cell, and the presence of a polarizer or analyzer. Usually $\gamma$ takes a value between 1.5 and 2.5 \citep{MacKintosh1989,Zhu1991}. In order to determine it precisely, we proceed to a DWS measurement without fluid motion ($\Omega=\SI{0}{\rpm}$). In this case, the decorrelation is only due to the Brownian motion of the scatterers : $\Gamma=0$, $1/\tau_v=0$ and $\langle \Delta \Phi_s(\tau)^2\rangle_s=4 \frac{\tau}{\tau_0} \frac{s}{l^*}$. Hence $g_1$ reduces to : 
\begin{equation}
g_1(\tau)=\exp\left(-\gamma\sqrt{6\tau/\tau_o}\right)
\label{g1BrownianMotion}
\end{equation}
and $\gamma$ can be determined since $\tau_0=1/(Dk^2)$ is now known for both fluids. In our setup, we find the following values of $\gamma$ : 2.27 (camera) and 2.31 (PMT) for the \ch{TiO2} suspension, 1.63 (camera) and 1.66 (PMT) for the lipid emulsion (see Table \ref{Tab:VariablesFluids}). Note that for the camera, the value of $\gamma$ slightly differs for each pixel. However, it is a narrow Gaussian distribution around the mean value $\overline{\gamma}$ (relative standard deviation $\sigma/\overline{\gamma}\approx5\%$), so we choose to perform the calculations with the same $\gamma=\overline{\gamma}$ for all pixels. This value can be considered as equal to the one found with the PMT, with less than $2\%$ difference. The difference between the fluid suspensions is due both to the different refractive indices and to the different geometries of the cells.

\subsubsection{Determination of $l^*$}
\label{l*}

The optical properties of Intralipid 20$\%$ have already been studied \citep{Michels2008}. At $\lambda=\SI{532}{\nano\metre}$ a scattering coefficient $\mu=\sigma c$ of about \SI{110}{\per\milli\metre}, corresponding to a mean free path $l=\SI{9.1}{\micro\metre}$, and an anisotropy factor $g=\langle \cos{\theta}\rangle$ of about 0.74, are found. This leads to a value of the transport mean free path of $l^*\approx  \SI{35}{\micro\metre}$ (see Table \ref{Tab:VariablesFluids}). For the \ch{TiO2} suspension, to our knowledge no precise data is available for the optical properties of Kronos 1002 dispersed in deionised water. Therefore we performed DWS measurements with a \SI{2}{\milli\metre} thick slab at different concentrations, in order to determine $l^*$ directly. Indeed, when a finite slab is used, the correlation function $g_1$ depends on the ratio $L/l^*$ (see appendix \ref{Appendix1}). We varied the concentration from \SI{0.5}{\gram/\liter} to \SI{5}{\gram/\liter} and found a linear relationship with the parameter $L/l^*$, as expected. We can therefore deduce that at a concentration of \SI{10}{\gram/\liter}, the transport mean free path is about $\SI{82}{\micro\metre}$ (see Table \ref{Tab:VariablesFluids}). It is difficult to compare this value to theoretical estimations, since the refractive index of anatase crystal is not very well known. If one assumes that it is close to the refractive index of rutile ($2.7\le n\le3$ for our wavelength), then one can deduce $l^*$ from Mie scattering theory. Such computation can be provided by https://omlc.org/calc/mie\_calc.html and one gets $\SI{36}{\micro\metre}\le l^*\le \SI{101}{\micro\metre}$ at our reference concentration of \SI{10}{\gram/\liter}, since anatase relative density is about 3.8. The mean free path is always bigger than \SI{20}{\micro\m}, thus $l\gg \lambda$ for both turbid fluids. Moreover, the transport mean free path $l^*$ for both fluids is much smaller than the thickness of the cell $L$, the characteristic length scale of the strain-rate tensor $\Lambda$ ($\Lambda\sim L \gg l^*$) or the pixel size.

\subsubsection{Measurement of the fluids viscosity}
\label{viscosity}

Since we use a rheometer head to drive the inner cylinder, we can extract the torque applied to the inner cylinder for a given rotation rate. In the linear Taylor-Couette flow, the link between the torque and the rotation rate via the dynamic viscosity for a Newtonian fluid is well-known \citep{Guyon2001_Book} : $T_i=4\pi\eta H\Omega R_o^2 R_i^2/(R_o^2-R_i^2)$. Therefore we can directly measure the viscosity of our turbid fluids by extracting the torque for different rotation rates in the linear regime. Note that the linear response of both fluids ensures their Newtonian behaviors. We find $\eta_{lipid}=\SI{2.63e-3}{\pascal\second}$ for the lipid emulsion and $\eta_{TiO_2}=\SI{0.86e-3}{\pascal\second}$ for the \ch{TiO2} suspension (close to the viscosity of water at 27°C). For both fluids, the density is very close (less than 1\% difference) to the density of water at 27°C : $\rho=\SI{997}{\kg/\m^3}$. The kinematic viscosity is therefore $\nu_{lipid}=\SI{2.64e-6}{\m^2/\s}$ for the lipid emulsion and $\nu_{TiO_2}=\SI{0.86e-6}{\m^2/\s}$ for the \ch{TiO2} suspension. The values of the Brownian motion correlation time $\tau_0$, the dimensionless coefficient $\gamma$, the transport mean free path $l^*$ and the kinematic viscosity $\nu$ of each fluid are shown in Table \ref{Tab:VariablesFluids} :

\vspace{0.5cm}
\begin{table}[!h]
	\centering
	\begin{tabular}{|c|c|c|}
		\hline
		  Variables & \ch{TiO2} suspension & Lipid emulsion \\[3pt]
		\hline
		  $\tau_0$ & \SI{3.39e-3}{\s} & \SI{2.37e-3}{\s} \\[3pt]
        \hline
            $\gamma$ (camera) & 2.27 & 1.63 \\[3pt]
		\hline
            $\gamma$ (PMT) & 2.31 & 1.66 \\[3pt]
        \hline
            $l^*$ & \SI{82}{\micro\m} & \SI{35}{\micro\m} \\[3pt]
        \hline
            $\nu$ & \SI{0.86e-6}{\m^2/\s} & \SI{2.64e-6}{\m^2/\s} \\[3pt]
        \hline
	\end{tabular} 
	\caption{Values of the key variables for each fluid in our experiment}
	\label{Tab:VariablesFluids}
\end{table}
\vspace{-0.5cm}

\subsubsection{Experimental protocol}
\label{Protocol}

The experiment proceeds as follows : 
\begin{itemize}
\item First we prepare the fluids as already mentioned. We fill the cell and mix the fluid by rotating the inner cylinder at \SI{100}{rpm} during \SI{180}{s}, in order to get a uniform concentration and therefore a uniform transport mean free path in the fluid.
\item We wait \SI{2}{\min} before doing a Brownian motion measurement ($\Omega=\SI{0}{rpm}$), to determine $\gamma$ (see section \ref{gamma}).
\item Then we alternate high and low rotation speeds to ensure a good mixing throughout the experiment and prevent inhomogeneity and flocking. All measurements are started \SI{1}{\min} after the rotation speed is changed in order to ensure a steady regime. The torque is recorded by the rheometer every second. The speckle intensity in the far-field region is measured by the PMT and the correlation is calculated by the correlator in real time. The sampling time of the correlator is \SI{1.28}{\micro\s} and the correlation function is averaged over \SI{1}{\min}. Simultaneously, the speckle pattern at the boundary between the outer cylinder and the fluid is recorded by the high-speed camera.  The acquisition time of the camera is \SI{2.5}{\micro\s}. For stationary processes, like the Taylor-Couette flow or the Taylor vortex flow, we average the correlation function over 100 000 images, so \SI{0.25}{s}. For time-dependent processes (see section \ref{wavy}), the averaging is done over only 25 000 images so \SI{0.0625}{s}, much less than the rotation period. The duration of the full measurement by the camera is then \SI{2}{s}, corresponding to 800 000 images, the maximum which can be recorded. The temporal resolution of the evolving spatially-resolved map of $\Gamma$ is therefore 1/16$^{th}$ of a second. In the end, the overall measurement duration for a given rotation rate is about \SI{3}{\min}.
\item We end the measurement run with a second measurement of the Brownian motion to ensure that no changes of the fluid properties occurred during the run, by checking that $\gamma$ has not changed.
\end{itemize}

\subsection{Data analysis}
\label{analysis}

In the setup with the PMT, the correlator directly computes the correlation function $g_2$ for 2048 points. On the contrary, there is some data processing to extract it from the camera images. Only 100 points of the correlation function are computed to speed up the averaging process, so that the correlation function $g_{2,p}$ of the pixel $p$ is given by :
\begin{equation}
g_{2,p}(m\Delta t)=\frac{\frac{1}{N-m} \sum\limits_{i=0}^{N-1-m} I_p(i\Delta t)I_p\left((i+m)\Delta t\right)}{\left(\frac{1}{N} \sum\limits_{i=0}^{N-1}I_p \left(i\Delta t\right)\right)^2}
\label{g2CAM}
\end{equation}
where $0\leq m\leq 99$, $N=100~000$ for stationary flows and $N=25~000$ for time-dependent flows.

\begin{figure}[h!]
\includegraphics[width=1\columnwidth,angle=0]{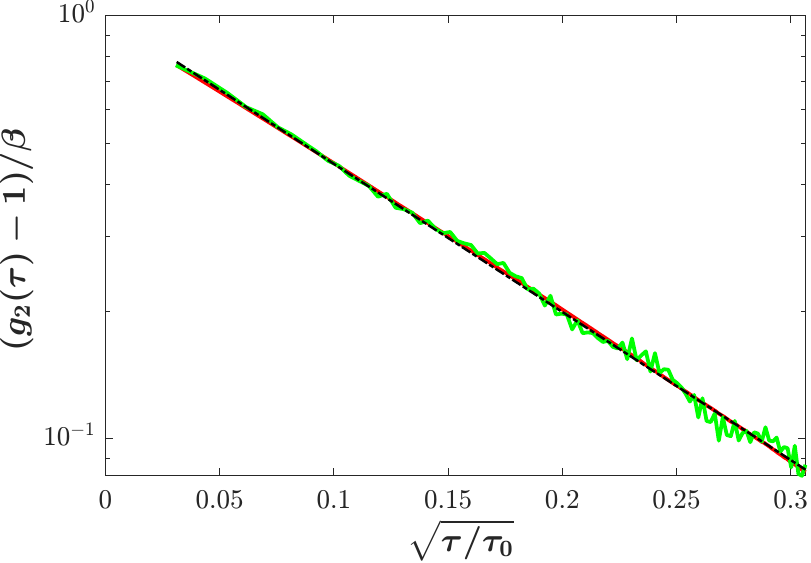}

\vspace{0.5cm}

\includegraphics[width=1\columnwidth,angle=0]{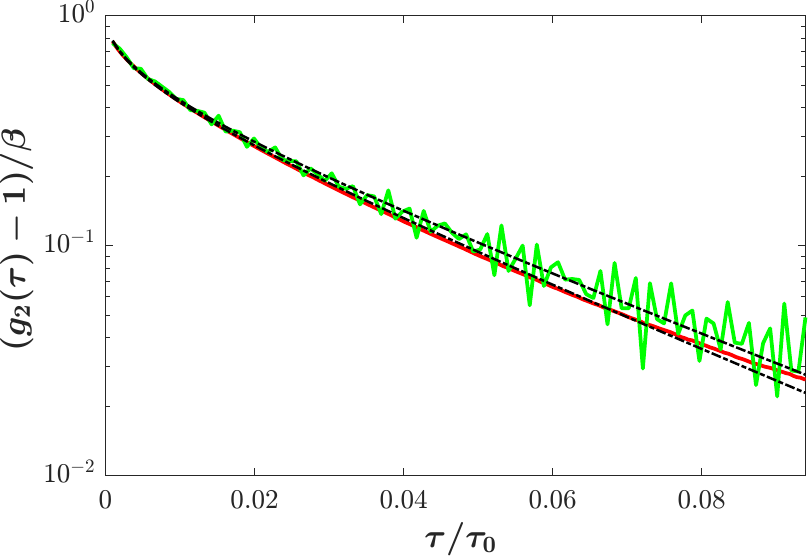} 
\caption{Normalised correlation function $(g_2-1)/\beta$ of the intensity of a speckle measured with the photomultiplier tube (in red)  and of a pixel with the high-speed camera (in green), {\bf Top:} when only Brownian motion is at play ($\Omega=\SI{0}{\rpm}$, therefore $1/\tau_v=\SI{0}{\per\s}$) and {\bf Bottom:} when a uniform shear is applied ($\Omega=\SI{30}{\rpm}$, $Ta=424$). The black dash--dotted lines correspond to the fit based on equation (\ref{g2exp}) for the PMT and the high-speed camera respectively.}
\label{SelfCorr}
\end{figure}

For each pixel correlation function, and for the PMT correlation function, we then remove the first points : 1 for the pixel, 2 for the PMT to start the correlation fits approximately at the same time (\SI{2.5}{\micro\s} vs \SI{2.56}{\micro\s}). This is done to avoid finite-size effects and very long trajectories probing the velocity gradients far from the surface, which may modify the exponential decay (see appendix \ref{Appendix1}). To fit, we keep the 99 remaining points of the pixels correlation function, i.e. up to $t=\SI{250}{\micro\s}$, and the corresponding 193 points of the PMT correlation function. In the end, we can fit the correlation function $g_2$ by its theoretical expression, following equations \eqref{Siegert} and \eqref{g1exp} : 

\begin{equation}
g_2(\tau)=\beta\exp\left(-2\gamma\sqrt{6\left [\tau/\tau_o+(\tau/\tau_v)^2\right]}\right)+1
\label{g2exp}
\end{equation}
where the contrast $\beta$ is obtained by extrapolating to $t=0$ : $\beta=g_2(0)-1$. Knowing $\gamma$ and $\tau_0$, we can deduce the correlation time $\tau_v$ associated with the velocity gradient and obtain $\Gamma=\frac{\sqrt{30}}{kl^*\tau_v}$. Figure \ref{SelfCorr} presents the normalized correlation function $(g_2-1)/\beta$ and shows that expression \eqref{g2exp} fits perfectly the measurements of both the PMT and the camera with a coefficient of determination higher than 0.99 for any pixel and higher than 0.995 for the PMT measurement. The curves are from the lipid emulsion setup, results of similar quality are obtained with the \ch{TiO2} suspension setup.

To get a global measurement from the camera and find the same results as with the PMT, we just need to compute the average of $\Gamma$ over all pixels. 


\section{Experimental results}
\label{results}

\subsection{Global measurements}
\label{cal}

The first step to validate the technique is to compare the DWS measurements to the theoretical prediction. In the circular Couette regime, the theoretical prediction for $\Gamma$ reduces to the shear rate at radius $R_o$ given by :
\begin{equation}
\Gamma_{th}(R_o)=\frac{2\Omega R_i^2}{R_o^2-R_i^2}
\label{Gammath}
\end{equation}

\begin{figure}[h!]
\includegraphics[width=1\columnwidth,angle=0]{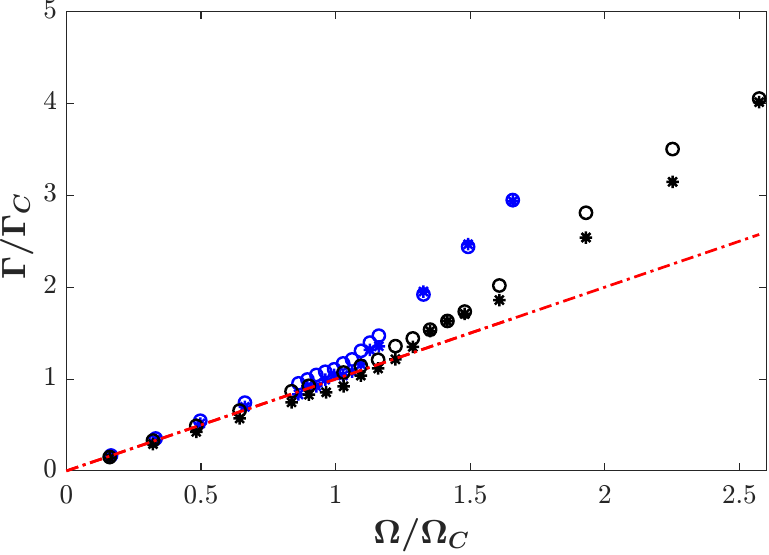} 
\caption{Scaled value of $\Gamma$ in the vicinity of the outer cylinder, measured with the PMT (circles) and the camera (stars) as a function of the scaled rotation rate. The data in blue corresponds to the lipid emulsion ($\Omega_c=\SI{60.3}{\rpm}$, $\Gamma_c=\SI{25.7}{\per\s}$) and the data in black corresponds to the \ch{TiO2} suspension ($\Omega_c=\SI{31.1}{rpm}$, $\Gamma_c=\SI{18.5}{\per\s}$). The red dash--dotted line represents the theoretical prediction of the laminar base flow.}
\label{meanShear}
\end{figure}

Figure \ref{meanShear} shows excellent agreement in the circular Couette regime between the theoretical prediction and the DWS measurements, both in the far-field region with the PMT and in the near-field region with the camera. Moreover, the discrepancy between the theory in the linear regime and the measurements appears close to the expected value of $Ta_c=1712$ corresponding to the first instability (Taylor vortex flow). It corresponds to a critical value of the rotation rate of $\Omega_c=\SI{60.3}{\rpm}$ for the lipid emulsion setup and $\Omega_c=\SI{31.1}{rpm}$ for the \ch{TiO2} suspension setup. The theoretical critical value of $\Gamma$, given by $\Gamma_c=2\Omega_c R_i^2/(R_o^2-R_i^2)$, is therefore $\Gamma_c=\SI{25.7}{\per\s}$ for the lipid emulsion setup and $\Gamma_c=\SI{18.5}{\per\s}$ for the \ch{TiO2} suspension setup.

\subsection{Spatially and temporally resolved measurements}
\label{resolved}

\subsubsection{Spatially resolved measurements}
\label{TCR}
\begin{figure}[!ht]
    \begin{subfigure}{0.31\columnwidth}
        \includegraphics[height=4.7cm]{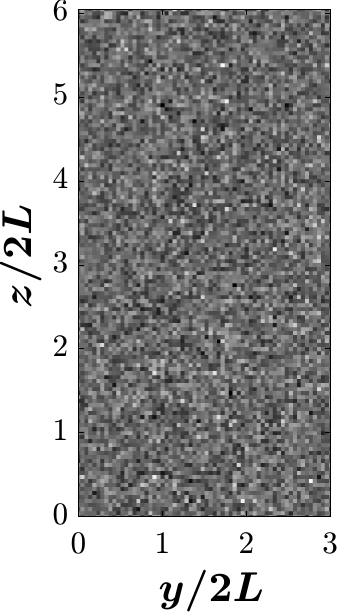}
    \end{subfigure}
    \begin{subfigure}{0.25\columnwidth}
        \includegraphics[height=4.7cm]{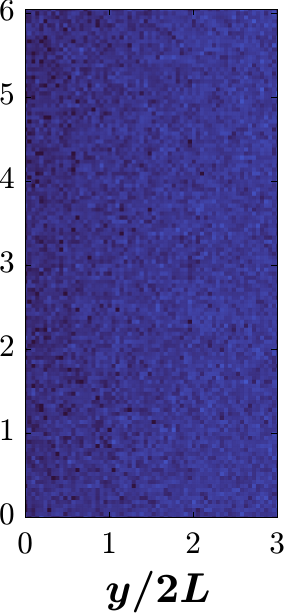}
    \end{subfigure}
    \begin{subfigure}{0.405\columnwidth}
        \includegraphics[height=4.7cm]{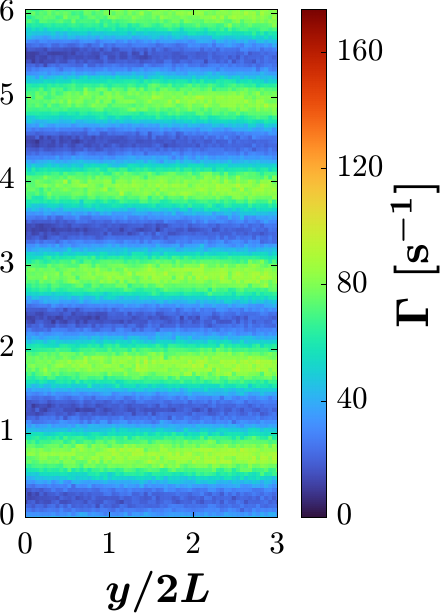}
    \end{subfigure}

    \caption{{\bf Left} Snapshot of the speckle pattern directly measured by the camera (arbitrary units of intensity). {\bf Middle} Spatially resolved map of $\Gamma$ in the linear regime ($\Omega=\SI{20}{\rpm}$, $Ta=188$ $<1712=Ta_c$), and {\bf Right} in the Taylor vortex regime ($\Omega=\SI{80}{\rpm}$, $Ta=3012$ $>Ta_c$}) with the lipid emulsion setup. The colorbar maximum (\SI{175}{\per\s}) corresponds to the maximum measured during the whole experiment, for $\Omega=\SI{100}{\rpm}$ (see Figure \ref{TemporalResolution}). The wavelength of the periodic pattern is about $2L$.
    \label{SpatialResolution}
\end{figure}

Up to now we have recovered the results of \cite{Bicout1994} with the PMT and have showed that a high-speed camera can also be used to get global measurements. We are additionally able to map $\Gamma$ at the surface. Since the area of measurement along the horizontal direction is small ($\approx \SI{1.6}{\cm}$) compared to the diameter of the outer cylinder, no curvature effect is observed and and we do not need to apply any correction to the images. Figure \ref{SpatialResolution} shows the spatially resolved maps of $\Gamma$ for different rotation speeds, with the lipid emulsion setup. Maps of similar quality are obtained with the \ch{TiO2} suspension setup. We can observe the homogeneous shear rate at the surface in the linear regime and the inhomogeneity of the norm of the strain-rate tensor at the surface in the Taylor vortex regime. Because of the Taylor vortices, $\Gamma$ exhibits a periodic behaviour with a wavelength of about $2 L$ at $Ta=3012$, as expected \citep{Bicout1994}.

\subsubsection{Spatio-temporal measurements in the wavy vortex regime}
\label{wavy}

\begin{figure}[h!]
\centering
\begin{subfigure}{0.31\columnwidth}
        \includegraphics[height=4.7cm]{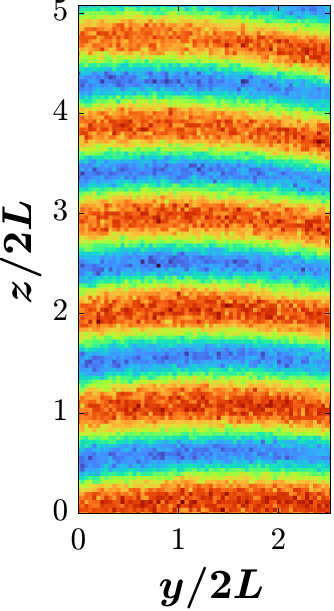}
    \end{subfigure}
    \begin{subfigure}{0.25\columnwidth}
        \includegraphics[height=4.7cm]{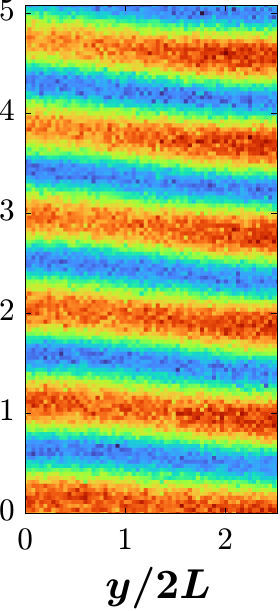}
    \end{subfigure}
    \begin{subfigure}{0.405\columnwidth}
        \includegraphics[height=4.7cm]{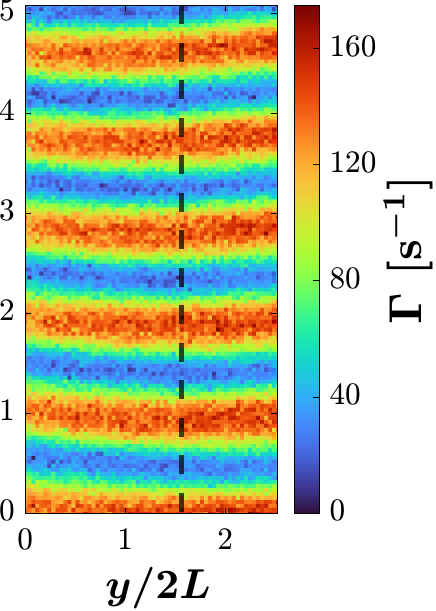}
    \end{subfigure}

\vspace{0.5cm}
    
    \begin{subfigure}{1\columnwidth}
        \includegraphics[height=4.7cm]{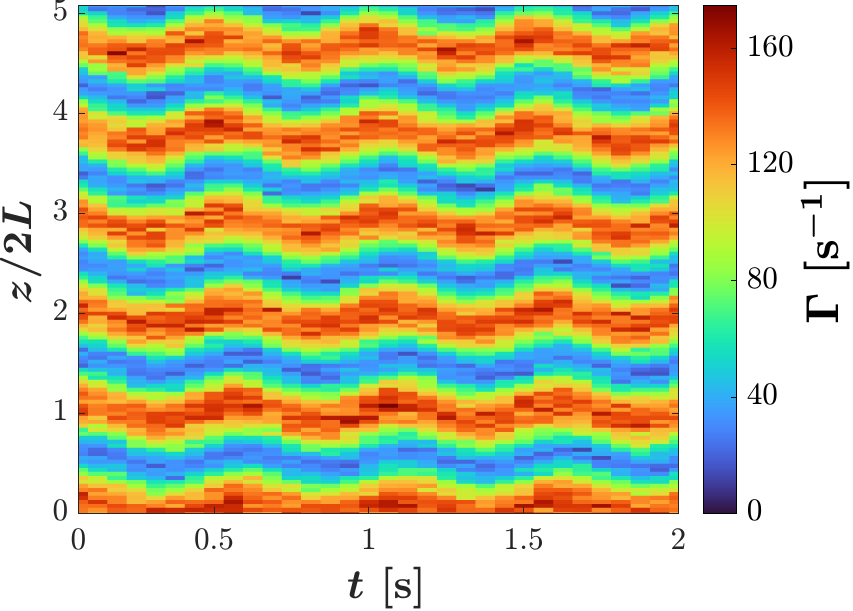}
    \end{subfigure}
\caption{{\bf Top} Three spatially resolved maps of $\Gamma$ (\SI{0.125}{\s} apart) with the lipid emulsion setup, in the wavy vortex regime ($\Omega=\SI{100}{\rpm}$, $Ta=4706$) where the oscillations of the vortices are clearly visible. {\bf Bottom} Time evolution of $\Gamma$ along a vertical line of pixels depicted by the black dashed line in the right map on top. The oscillation period is about \SI{0.52}{\s}.}
\label{TemporalResolution}
\end{figure}

By averaging the correlation function over 25 000 images, we are able to map the norm of the strain-rate tensor with a period of \SI{0.0625}{s} (1/16$^{th}$ of a second). Therefore, we are able to highlight the oscillations of the vortices observed in the wavy vortex regime. Figure \ref{TemporalResolution} presents this spatio-temporal resolved measurement for $\Omega=\SI{100}{\rpm}$ ($Ta=4706$ for the lipid emulsion). The three spatially resolved maps at the top exhibit different orientations of the vortices \SI{0.125}{\s} apart. They correspond to three different phases of the wavy motion of the vortices, illustrated for a column of pixels in the temporal evolution diagram at the bottom. From this diagram we can extract the oscillation period (about $\SI{0.52}{\s}$). Since there are 4 azimuthal waves, the wave speed is about 0.29 $\Omega$, which is consistent with previously reported values \citep{King1984}. An animation of the complete measurement (\SI{2}{\s}) is available in the supplementary material.

\section{Conclusions}
\label{conclu}

This work presents an optical technique, Diffusing-Wave Spectroscopy, to measure directly the norm of the strain-rate tensor and thus the energy density dissipation rate in a Newtonian fluid. The main advantage of this technique is that it does not necessitate the spatial differentiation of a velocity measurement to measure the dissipation rate. Moreover, velocity gradients are probed on a very small length scale : the transport mean free path $l^*$, so about 10 to \SI{100}{\micro\m}. Our new input is the use of a high-speed camera that allows us to get a measurement of the norm of the strain-rate tensor resolved in space and time. We apply this novel technique to the well-known Taylor-Couette flow to test its accuracy, and we show that the method is quantitative. It enables us to get a time-dependent map of the norm of the strain-rate tensor, from the circular Couette regime up to the wavy vortex regime. This technique still has some limitations. In the backscattering geometry, the measurement is restricted to the vicinity of a boundary. So far, in our case, the resolution of the camera must be reduced to a frame of 64$\times$128 pixels to reach a sufficiently high frame rate to measure the decay of the correlation functions. The time resolution is also limited to 1/16 \SI{}{\s} by the convergence of the correlation functions. This last point might be further optimized and the ever-increasing performance of high-speed cameras should help to overcome these limitations. Having proved the concept of this technique, it may now be applied to flows where knowledge of the dissipation rate is particularly relevant, for instance around structures immersed in turbulent flow. Indeed, with a big enough experiment, the Kolmogorov scale can be significantly bigger than $l^*$, enabling the measurement of the velocity gradient at sufficiently small scales. If we consider the lipid emulsion with a transport mean free path of \SI{35}{\micro\m}, and a characteristic length scale of \SI{1}{m} for the energy injection, a DWS measurement should be able to properly measure the dissipation rate for a typical Kolmogorov scale as small as $5l^*=\SI{175}{\micro\m}$, so up to a Reynolds number of $10^5$. Moreover, this would correspond to a typical value of $\Gamma$ of \SI{83}{\s^{-1}}, below the critical value of \SI{200}{\s^{-1}}. Lastly,  by varying the pixels field of view by zooming in or out, the area over which the dissipation rate is averaged can be modified, enabling a wide range of wave numbers to be explored. We hope that this work will be a  helpful starting point to  design the setup dedicated to such studies.

\appendix*
\section{Boundary conditions and finite-size effects}
\label{Appendix1}

To exactly compute the correlation function $g_1$, one needs to solve the diffusion equation to determine the probability density of paths length $P$ \citep{Weitz1993_Chapter,Sheng2006_Chapter}. To do so, we have to choose the initial and boundary conditions (BC) describing the diffusive transport of the light in the cell. We consider a slice of thickness $L$ in the $x$ direction ($0\leq x\leq L$) and of infinite extent in the $y$ and $z$ directions. For the initial condition, in the case of uniform illumination on the incident face, the initial "diffusive light" (in the sense of being described by the diffusion equation) is often described in the DWS theory by a Dirac (with infinite extent in $y$ and $z$) at a distance $x_0$ from the incident face. Indeed, the transport of light can be described as diffusive only once the incident light has been scattered. We expect that the first scattering event happens at a distance of order $l^*$ from the incident face, so $x_0\approx l^*$. For the boundary conditions, we can decide to set the flux of diffusive light into the cell to zero at the boundaries, since no scattered light enters the sample from outside. It is even more relevant to set the flux of diffusive light into the cell to a fraction $R$ of the flux of diffusive light leaving the cell, to take into account reflections at the boundaries. This is the partial-current BC. An equivalent BC is the extrapolated BC : the density of diffusive light is set to 0 at an extrapolation length $C=\frac{2}{3}l^*\frac{1+R}{1-R}$ outside the cell \citep{Zhu1991,Haskell1994}. We found that this solution is in even better agreement with our experimental data than the partial-current BC. Other boundary conditions are possible, such as the absorbing BC, but they usually provide solutions less in agreement with experiments \citep{Weitz1993_Chapter,Pine1990}. The probability density of path lengths $P(s)$ and its Laplace transform (the correlation function $g_1(\tau)$) can be obtained from chapter 14.3 in \cite{Carslaw1959}. For the extrapolated BC, in backscattering (i.e. looking at the diffusive light at x=0), we obtain :


\begin{equation}
P(s)= \frac{\pi^2 l^*\sum\limits_{n=1}^{\infty}\sin\left(\frac{n\pi C}{L+2C}\right)\sin\left(\frac{n\pi x_0}{L+2C}\right)\exp\left(-\frac{n^2\pi^2 l^* s}{3(L+2C)^2}\right)}{3(L+2C)^2\sum\limits_{n=1}^{\infty}\frac{1}{n^2}\sin\left(\frac{n\pi C}{L+2C}\right)\sin\left(\frac{n\pi x_0}{L+2C}\right)}
\label{Eq:P(s)Extrapolated}
\end{equation}

\begin{equation}
g_1(\tau)= \frac{\sinh\left(\frac{C}{l^*}\sqrt{6T}\right)\sinh\left(\frac{L+C-x_0}{l^*}\sqrt{6T}\right)}{\left(1-\frac{x_0+C}{L+2C}\right)\frac{C}{l^*}\sqrt{6T}\sinh\left(\frac{L+2C}{l^*}\sqrt{6T}\right)}
\label{g1Extrapolated}
\end{equation}
where $T=\tau/\tau_0+\tau^2/\tau_v^2$. In the limit of a semi-infinite medium ($l^*\ll L$), the correlation function reduces to : 


\begin{equation}
g_1(\tau)= \frac{l^*}{C\sqrt{6T}}\exp(-\frac{x_0+C}{l^*}\sqrt{6T})\sinh(\frac{C}{l^*}\sqrt{6T})
\label{Eq:g1ExtrapolatedSemiInfini}
\end{equation}

In the limit of short times ($T\ll1$), the decay is almost exponential : 
\begin{equation}
g_1(\tau)\approx\exp(-\gamma \sqrt{6T})
\label{g1Simplified}
\end{equation}
where 
$\gamma=\frac{x_0+C}{l^*}=\frac{x_0}{l^*}+\frac{2}{3}\frac{1+R}{1-R}$. The dimensionless parameter $\gamma$ is therefore linked to $x_0$ but also to the geometry of the cell and the refractive indices of the fluid and the cell through $R$. It is also known to depend on the presence of a polarizer or analyzer, since these can foster shorter or longer paths \citep{Pine1990,Weitz1993_Chapter}. 

\begin{figure}[h!]
\includegraphics[width=1\columnwidth,angle=0]{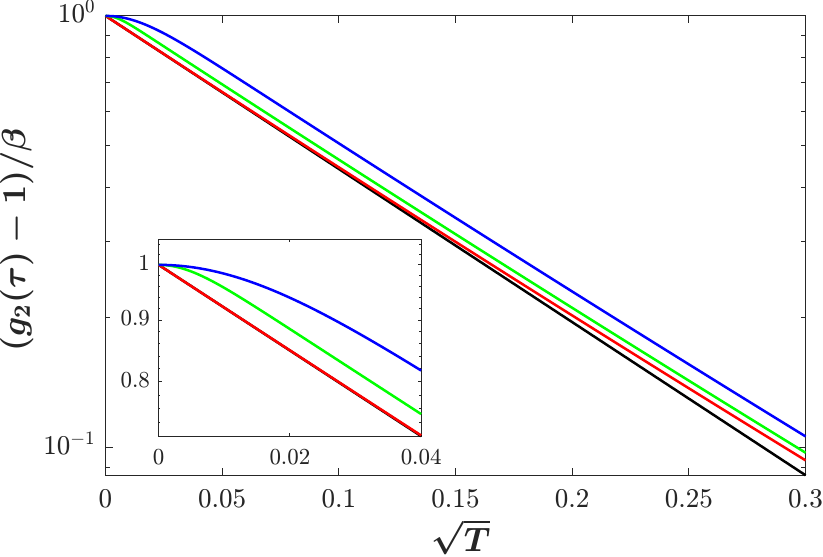} 
\caption{Normalised correlation function $(g_2(\tau)-1)/\beta=|g_1(\tau)|^2 $ of the intensity for $L/l^*=25$ (in blue) corresponding to the \ch{TiO2} suspension setup and $L/l^*=81$ (in green) corresponding to the lipid emulsion setup, from equation \eqref{g1Extrapolated}. The semi-infinite medium case (in red) from equation \eqref{Eq:g1ExtrapolatedSemiInfini} and the exponential approximation (in black) from equation \eqref{g1Simplified} are plot for comparison. The inset zooms in the early stage of the curve where the finite size effects are significant.}
\label{Fig:FiniteSizeEffects}
\end{figure}

Figure \ref{Fig:FiniteSizeEffects} highlights the finite size effects on the normalized correlation function of the intensity $(g_2(\tau)-1)/\beta=|g_1(\tau)|^2$, for $x_0=l^*$, $C=2l^*/3$ ($R=0$, no reflection) and the corresponding $\gamma=5/3$. When the ratio $L/l^*$ decreases, deviation from the exponential behaviour is observed at very short times. It corresponds to a reduction of the contribution of very long paths, since they can be transmitted and therefore lost for backscattering. To avoid these effects, we remove the very first points in our correlation functions and extrapolate the initial value $\beta=g_2(0)-1$ (see section \ref{analysis}). Note that at slightly longer times, the slopes are the same and are very close to the exponential approximation. We can also focus on these differences at very short time to measure $l^*$ by fitting equation \eqref{g1Extrapolated} to the experimental data, as long as $x_0$ and $C$ are known from a "semi-infinite" measurement fitted with equation \eqref{Eq:g1ExtrapolatedSemiInfini}.

\section*{Declarations}

\begin{itemize}
\item Acknowledgements :\\
The authors would like to thank Jérôme Crassous for introducing them to DWS, Vincent Padilla for helping them build the setup, Patrick Guenoun for giving them access to DLS facilities and Kronos\textsuperscript{TM} for providing a free sample of \ch{TiO2} particles. We are grateful to Basile Gallet, Christopher Higgins, Fabrice Charra, \correct{Michael Berhanu,} Aliz\'ee Dubois, Marco Bonetti and Dominique Bicout for insightful discussions.

\item Ethical Approval:\\
No applicable.
 
\item Competing interests:\\
There is no competing interests
 
\item Authors' contributions:\\
E.F. and S.A. participated equally at all stages of this work and wrote the main manuscript text. V.B. and T.W. participated to the early development of the experiment. All authors reviewed the manuscript.

\item Funding:\\
This research is supported by the French National Research Agency (ANR DYSTURB Project No. ANR-17-CE30-0004) and the European Research Council under grant agreement (project FLAVE 757239).
 
\item Availability of data and materials:\\
All datasets are available on request from the corresponding author
\end{itemize}

\bibliography{Bibliographie}

\end{document}